\pdfoutput=1
\documentclass[10pt]{iopart}

\usepackage{iopams}
\expandafter\let\csname equation*\endcsname\relax
\expandafter\let\csname endequation*\endcsname\relax

\usepackage{graphicx}
\usepackage{amsmath, amssymb, mathrsfs, amsthm}
\usepackage{color}
\usepackage{bm}
\usepackage[T1]{fontenc}
\usepackage[utf8]{inputenc}
\usepackage{microtype}
\usepackage[normalem]{ulem}
\usepackage[dvipsnames]{xcolor}
\usepackage[colorlinks,urlcolor=NavyBlue,citecolor=NavyBlue,linkcolor=NavyBlue,pdfusetitle]{hyperref}
\usepackage[all]{hypcap}
\usepackage{orcidlink}
\usepackage{verbatim}

\graphicspath{{figs/}}

\newcommand{\lcs}[1]{\textcolor{WildStrawberry}{#1}}

\newcommand{\pd}{\partial}
\newcommand{\cd}{\nabla}

\begin{document}

\note{Can a radiation gauge be horizon-locking?}

\author{Leo C.\ Stein\,\orcidlink{0000-0001-7559-9597}}
\ead{lcstein@olemiss.edu}
\address{Department of Physics and Astronomy,
    University of Mississippi, University, MS 38677, USA}

\hypersetup{pdfauthor={Stein},
  pdftitle={Can a radiation gauge be horizon-locking?}}

\date{\today}

\begin{abstract}
  In this short Note, I answer the titular question: Yes, a radiation
  gauge can be horizon-locking.  Radiation gauges are very common in
  black hole perturbation theory.  It's also very convenient if a
  gauge choice is horizon-locking, i.e.~the location of the horizon
  is not moved by a linear metric perturbation.  Therefore it is
  doubly convenient that a radiation gauge can be horizon-locking,
  when some simple criteria are satisfied.  Though the calculation is
  straightforward, it seemed useful enough to warrant writing this
  Note.  Finally I show an example: the $\ell$ vector of the
  Hartle--Hawking tetrad in Kerr satisfies all the conditions for
  ingoing radiation gauge to keep the future horizon fixed.
\end{abstract}

The context of this Note is black hole perturbation theory
(see~\cite{Pound:2021qin} for a review).  Suppose we
have a Lorentzian spacetime $(M,\mathring{g})$ where $\mathring{g}$ is
the background metric, e.g. the Kerr metric
(see~\cite{Teukolsky:2014vca} for a review), which has a future
horizon $\mathcal{H}^{+}$ (see~\cite{Poisson:2009pwt} for a
pedagogical introduction).  We work to first order
in perturbation theory, with a metric
\begin{align}
  \label{eq:pert-g}
  g_{ab} = \mathring{g}_{ab} + \varepsilon h_{ab} + \mathcal{O}(\varepsilon^{2})
  \,,
\end{align}
where $\varepsilon$ is a formal order-counting parameter.

Chrzanowski introduced two ``radiation gauges'' for perturbations
in~\cite{Chrzanowski:1975wv}.  These radiation gauges are adapted for
algebraically special~\cite{Stewart:1990uf} spacetimes.  If $\ell^{a}$
is an \emph{outgoing} principal null vector field, then \emph{ingoing}
radiation gauge (IRG) is specified by
\begin{align}
  \label{eq:l-dot-h}
  \ell^{a}h_{ab} =0 \,, \quad h\equiv\mathring{g}^{ab}h_{ab} = 0
  \qquad\text{(IRG)} \,.
\end{align}
Similarly, if $n^{a}$ is an ingoing principal null vector field, then
outgoing radiation gauge (ORG) is the same but with $n$ replacing
$\ell$.  These gauges at first seem over-specified, with 5 algebraic
conditions.  Price et al.~\cite{Price:2006ke} showed that one of IRG
or ORG is admissible in a Petrov type II metric, whereas in type D,
both are admissible.

The event horizon is the defining feature of a black
hole~\cite{Wald:1984rg}, and thus it is of great physical interest to
locate the horizon, e.g.~to study
thermodynamics~\cite{Hollands:2024vbe} or
tides~\cite{OSullivan:2015lni}, or to compute fluxes down the
horizon~\cite{Hawking:1972hy}.
In general, locating a horizon is challenging since it is
teleological, requiring global knowledge of the entire future
development of the spacetime~\cite{Wald:1984rg}.
This challenge is lessened in perturbation theory, but replaced with
the new challenge that we are free to make $\mathcal{O}(\varepsilon)$
coordinate transformations.  These generate the gauge transformations
$h_{ab} \to h_{ab}' = h_{ab} + \mathcal{L}_{\xi} \mathring{g}_{ab} =
h_{ab} + \mathring{\cd}_{(a}\xi_{b)}$ where $\xi^{a}$ generates the
infinitesimal diffeomorphism.  We are describing the same physical
spacetime, but the horizon moves by $\mathcal{O}(\varepsilon)$ in
coordinates.

On the other hand, we can exploit this freedom to make coordinates of
the horizon of $g_{ab}$ coincide with the analytically-known horizon
of $\mathring{g}_{ab}$.  A gauge choice achieving this is called
``horizon-locking,'' possibly introduced by~\cite{Seidel:1992vd},
though the idea is surely older.  There is still considerable freedom
in achieving a horizon-locking gauge: only components of $\xi^{a}$
transverse to the horizon are relevant~\cite{Poisson:2004cw}.
We can now pose the question asked in the title of this Note: Can a
radiation gauge be horizon-locking?  Yes.

\newtheorem*{thm}{Theorem}
\begin{thm}
  Let $(M,\mathring{g})$ be a stationary, Ricci-flat, Lorentzian
  spacetime with future horizon $\mathcal{H}^{+}$.  Let $\ell^{a}$:
  (i)~be null, %
  (ii)~be geodesic, and %
  (iii)~generate $\mathcal{H}^{+}$.
  Let $h_{ab}$ be the perturbation as in Eq.~\eqref{eq:pert-g}, and
  let $h_{ab}$ vanish either in the distant past or future.  Further
  let $R_{ab}=\mathcal{O}(\varepsilon^{2})$ with $R_{ab}$ the Ricci
  tensor of $g_{ab}$.  Then the gauge Eq.~\eqref{eq:l-dot-h} is
  horizon-locking.
\end{thm}
\begin{proof}
  First, we follow~\cite{Poisson:2004cw} to see that the event horizon
  and apparent horizon agree to first order in $\varepsilon$.  Consider
  the Raychaudhuri equation for a geodesic
  null congruence $k^{a}$ that generates the horizon,
  with affine parameter $v$,
  \begin{align}
    \label{eq:focusing}
    \frac{d\theta}{dv} = -\frac{1}{2} \theta^{2} - \sigma_{ab} \sigma^{ab}
    + \omega_{ab}\omega^{ab} - R_{ab}k^{a}k^{b}
    \,.
  \end{align}
  Here $\theta$ is the expansion scalar, $\sigma_{ab}$ is the shear,
  and $\omega_{ab}$ is the twist.  By assumption, the Ricci term
  vanishes at zeroth and first order.  Since the horizon
  generator is hypersurface orthogonal,
  $\mathring{\omega}_{ab}|_{\mathcal{H}^{+}}=0=\omega_{ab}|_{\mathcal{H}^{+}}$.

  Expand all quantities as a series in $\varepsilon$,
  e.g.~$\sigma_{ab} = \mathring{\sigma}_{ab} + \varepsilon
  \sigma^{(1)}_{ab} + \mathcal{O}(\varepsilon^{2})$.  Stationarity of the
  background then tells us that
  $d\mathring{\theta}/dv|_{\mathcal{H}^{+}}=0$, and thus
  $\mathring{\theta}|_{\mathcal{H}^{+}}=\mathring{\sigma}_{ab}|_{\mathcal{H}^{+}}=0$.
  Now study the $\mathcal{O}(\varepsilon^{1})$ equation, which says
  \begin{align}
    \frac{d\theta^{(1)}}{dv} = - \mathring{\theta}\theta^{(1)}
    - 2\mathring{\sigma}^{ab} \sigma^{(1)}_{ab}
    + 2\mathring{\omega}^{ab}\omega^{(1)}_{ab}
    \,.
  \end{align}
  Evaluating at the background horizon, all terms on the right-hand
  side vanish, so $\theta^{(1)}|_{\mathcal{H}^{+}}$ is constant.
  Since $h_{ab}$ vanishes in the distant past or future, this constant
  must be $\theta^{(1)}|_{\mathcal{H}^{+}} = 0$.  Therefore the
  perturbed event horizon is an apparent horizon to
  $\mathcal{O}(\varepsilon^{1})$, and our job has reduced to locating the
  apparent horizon at first order.

  Now for locating the apparent horizon.  First note that in the gauge
  \eqref{eq:l-dot-h}, the vector field $\ell^{a}$ is automatically
  null up to our desired order,
  \begin{align}
    g(\ell,\ell) = \mathring{g}_{ab}\ell^{a}\ell^{b} + \varepsilon h_{ab}\ell^{a}\ell^{b} + \mathcal{O}(\varepsilon^{2}) = 0 + \varepsilon 0 + \mathcal{O}(\varepsilon^{2}) \,.
  \end{align}
  Correspondingly, lowering $\ell^{a}$ with either metric
  gives the same one-form, $\ell_{a}\equiv g_{ab}\ell^{b} =
  \mathring{g}_{ab} \ell^{b} + O(\varepsilon^{2})$.  Therefore we find no
  need to expand $\ell^{a}$ in a series in $\varepsilon$.
  Below we need an identity arising from a gradient of the gauge
  conditions \eqref{eq:l-dot-h},
  \begin{align}
    \label{eq:exchange-l-h-derivs}
    \mathring{\cd}_{a}(\ell^{c}h_{cd}) = 0
    \qquad \Rightarrow \qquad
    \ell^{c}\mathring{\cd}_{a}h_{cd} = -h_{cd}\mathring{\cd}_{a}\ell^{c} \,.
  \end{align}
  Here $\mathring{\cd}$ is the Levi-Civita connection of
  $\mathring{g}$.
  
  Now let's check that $\ell^{a}$ is geodesic with respect to the
  perturbed metric, not just the background metric.  To do this we
  need to express the
  Levi-Civita connection of $g$, which we call $\cd$, in terms of
  $\mathring{\cd}$.  The two connections are related by
  \begin{align}
    \label{eq:cd-minus-cd0}
    \cd_{b} v^{a} - \mathring{\cd}_{b} v^{a} = \varepsilon \dot{C}^{a}{}_{bc} v^{c} + \mathcal{O}(\varepsilon^{2}) \,,
  \end{align}
  where the linearized difference of connections tensor is~\cite{Wald:1984rg}
  \begin{align}
    \label{eq:dot-C}
    \dot{C}^{a}{}_{bc} = \frac{1}{2} \mathring{g}^{ad}
    \left[
      \mathring{\cd}_{b}h_{cd} + \mathring{\cd}_{c}h_{bd}
      - \mathring{\cd}_{d}h_{bc}
    \right]
    \,.
  \end{align}
  By assumption, with the background connection we have a geodesic
  congruence, not affinely parameterized,
  \begin{align}
    \ell^{a}\mathring{\cd}_{a}\ell^{b} = \mathring{\kappa} \ell^{b} \,.
  \end{align}
  Evaluate $\cd_{\ell}\ell$ to see if it's geodesic:
  \begin{align}
    \ell^{a}\cd_{a} \ell^{b} &= \ell^{a}\mathring{\cd}_{a} \ell^{b} + \varepsilon \ell^{a} \dot{C}^{b}{}_{ac}\ell^{c} + \mathcal{O}(\varepsilon^{2}) \,, \\
    &= \mathring{\kappa} \ell^{b}+ \varepsilon \frac{1}{2}  \ell^{a}\ell^{c}
    \mathring{g}^{bd} 
    \left[
      \mathring{\cd}_{a}h_{cd} + \mathring{\cd}_{c}h_{ad}
      - \mathring{\cd}_{d}h_{ac}
    \right] + \mathcal{O}(\varepsilon^{2}) \,, \\
    &= \mathring{\kappa} \ell^{b}+ \varepsilon
    \mathring{g}^{bd}
    \left[
      \ell^{a}\ell^{c}\mathring{\cd}_{a}h_{cd}
      - \frac{1}{2} \ell^{a}\ell^{c}\mathring{\cd}_{d}h_{ac}
    \right] + \mathcal{O}(\varepsilon^{2}) \,, \\
    &= \mathring{\kappa} \ell^{b}+ \varepsilon
    \mathring{g}^{bd}
    \left[
      -\ell^{a}h_{cd}\mathring{\cd}_{a}\ell^{c}
      + \frac{1}{2} \ell^{a}h_{ac}\mathring{\cd}_{d}\ell^{c}
    \right] + \mathcal{O}(\varepsilon^{2}) \,, \\
     &= \mathring{\kappa} \ell^{b}+ \varepsilon 
    \mathring{g}^{bd}
    \left[
      -h_{cd}\mathring{\kappa}\ell^{c}
      + 0
    \right] + \mathcal{O}(\varepsilon^{2}) \,,\\
    \ell^{a}\cd_{a} \ell^{b} &= \mathring{\kappa} \ell^{b} + \mathcal{O}(\varepsilon^{2}) \,.
  \end{align}
  Therefore $\ell$ is also still a null geodesic congruence with
  respect to $g$, not just $\mathring{g}$.  Furthermore, the
  inaffinity has not changed,
  $\kappa=\mathring{\kappa} + \mathcal{O}(\varepsilon^{2})$, a result
  we need below.

  Next we need to check that $\ell^{a}$ is still
  hypersurface-orthogonal.  From the Frobenius theorem, the one-form
  $\ell_{a}$ is hypersurface-orthogonal when $\ell \wedge d\ell = 0$.
  This has implicit dependence on the metric, lowering the vector
  $\ell^{a}$ into the one-form.  As we saw above, the gauge condition
  makes
  $g_{ab}\ell^{b}=\mathring{g}_{ab}\ell^{b}+\mathcal{O}(\varepsilon^{2})$.
  Therefore whenever $\ell \wedge d\ell$ vanishes according to the
  background metric, it also vanishes according to the perturbed
  metric, up to $\mathcal{O}(\varepsilon^{2})$.  Thus $\ell_{a}$ is
  hypersurface-orthogonal at $\mathcal{H}^{+}$ with respect to both
  metrics.

  Finally we want to check that the congruence $\ell^{a}$ has
  vanishing expansion---as measured with $g_{ab}$---at the unperturbed
  horizon.  To find the expansion, we proceed as
  usual~\cite{Poisson:2009pwt} by studying
  $B_{ab} \equiv \cd_{b}\ell_{a}$.  Specifically we will need to take
  an orthogonal projection with the aid of an auxiliary null vector
  $n_{a}$, satisfying $n_{a}\ell^{a}=-1$ (we work in signature
  ${-}{+}{+}{+}$).
  Next construct the
  orthogonal projector $\gamma_{ab} = g_{ab} +\ell_{a}n_{b}+n_{a}\ell_{b}$,
  and use it to project out
  $\hat{B}_{ab} = \gamma_{a}{}^{c}\gamma_{b}{}^{d}B_{cd}$.  The
  expansion scalar is the trace,
  \begin{align}
    \theta &= \gamma^{ab} \hat{B}_{ab} = \gamma^{ab} B_{ab}
    = g^{ab} \cd_{b}\ell_{a} + \ell^{a}n^{b}\cd_{b}\ell_{a} + n^{a}\ell^{b}\cd_{b}\ell_{a} \,,\\
    \theta &= \cd_{a}\ell^{a} + n^{b}\cd_{b}(\tfrac{1}{2}\ell^{a}\ell_{a}) + n^{a} \kappa \ell_{a} = \cd_{a}\ell^{a} - \kappa \,.
  \end{align}
  In this final expression we see that all references to $B_{ab}$ and
  the auxiliary $n^{a}$ have disappeared, so we don't have to worry
  about their perturbations; we just need this last expression along with
  $\mathring{\theta} = \mathring{\cd}_{a}\ell^{a} - \mathring{\kappa}$.
  The perturbed expansion is
  \begin{align}
    \theta &= \cd_{a}\ell^{a} -\kappa
    = \mathring{\cd}_{a}\ell^{a} + \varepsilon \dot{C}^{a}{}_{ab}\ell^{b} -\mathring{\kappa} + \mathcal{O}(\varepsilon^{2}) \\
    &= \mathring{\theta} + \varepsilon \ell^{b}
    \frac{1}{2} \mathring{g}^{ad}
    \left[
      \mathring{\cd}_{a}h_{bd} + \mathring{\cd}_{b}h_{ad}
      - \mathring{\cd}_{d}h_{ab}
    \right] + \mathcal{O}(\varepsilon^{2}) \,,\\
    \theta &= \mathring{\theta} + \varepsilon \ell^{b}
    \frac{1}{2}
    \left[
      \mathring{\cd}_{a}h_{b}{}^{a} + \mathring{\cd}_{b}h
      - \mathring{\cd}_{d}h^{d}{}_{b}
    \right] + \mathcal{O}(\varepsilon^{2}) = \mathring{\theta} + \mathcal{O}(\varepsilon^{2}) \,.
  \end{align}
  The first and third term in parentheses cancel, and the middle term
  vanishes from the gauge condition for vanishing trace $h=0$.  Thus
  we have shown that the perturbed expansion is the same as the
  background expansion up to $\mathcal{O}(\varepsilon^{2})$.  In
  particular, $\theta$ (as measured by $\cd_{a}$) vanishes at the
  unperturbed horizon $\mathcal{H}^{+}$, thus locating the perturbed
  apparent horizon; which we saw above is the same as the perturbed
  event horizon.
\end{proof}
\theoremstyle{remark}
\newtheorem{rem}{Remark}
\begin{rem}
  Notice that the conditions for the theorem are weaker than what is
  usually done in black hole perturbation theory: $\ell^{a}$ does
  not need to be a principal null direction.
\end{rem}
\begin{rem}
  The condition $R_{ab}=\mathcal{O}(\varepsilon^{2})$ is satisfied if
  $h_{ab}$ solves the linearized Einstein equations with vanishing
  first-order source $T_{ab}$.  For example, in the EMRI problem we
  have a point-particle source, so
  $R_{ab}=\mathcal{O}(\varepsilon^{2})$ everywhere except the location
  of the particle.  Horizon-locking can be achieved at all times
  except when the particle passes through the horizon.
\end{rem}
\begin{rem}
  The condition $R_{ab}=\mathcal{O}(\varepsilon^{2})$ can be
  generalized to the weaker condition
  $R_{ab}\ell^{a}\ell^{b} = \mathcal{O}(\varepsilon^{2})$.
\end{rem}
\begin{rem}
  Throughout the derivation, we only needed the gauge condition
  \eqref{eq:l-dot-h} and its first derivative evaluated along
  $\mathcal{H}^{+}$.  Therefore, the theorem still holds replacing the
  global gauge condition with just the horizon boundary condition
  \begin{align}
    \ell^{a}h_{ab}\Big|_{\mathcal{H}^{+}} &=0 \,, \quad h\Big|_{\mathcal{H}^{+}} = 0 \,, \quad
    \mathring{\cd}_{c}(\ell^{a}h_{ab})\Big|_{\mathcal{H}^{+}} =0 \,, \quad \mathring{\cd}_{a}h\Big|_{\mathcal{H}^{+}} = 0 \,.
  \end{align}
\end{rem}
\begin{rem}
  Using $n^{a}$ and its ingoing expansion in place of $\ell^{a}$ and
  its outgoing expansion, and using $\mathcal{H}^{-}$ in place of
  $\mathcal{H}^{+}$, the theorem also applies to ORG being compatible
  with fixing the past horizon.
\end{rem}

\section*{Example: Hartle--Hawking tetrad for the Kerr metric}
\newcommand{\HH}{\text{HH}}
\newcommand{\np}{\text{NP}}
Here we show that the $\ell$ vector of the Hartle--Hawking tetrad
for the Kerr metric satisfies the conditions for the above theorem.
Our metric is compactly represented by specifying our
tetrad.  We use ingoing coordinates $(v,r,\theta,\tilde{\phi})$ to
give the Hartle--Hawking tetrad components~\cite{Hawking:1972hy,
  Teukolsky:1973ha, Teukolsky:1974yv},
\begin{align}
  \ell^{a} &= \left( 1, \frac{1}{2}\frac{\Delta}{r^{2}+a^{2}},0,\frac{a}{r^{2}+a^{2}}\right) \,,\\
  n^{a} &= \left( 0, -\frac{r^{2}+a^{2}}{\Sigma}, 0, 0\right) \,,\\
  m^{a} &= \frac{1}{2(r+ia\cos\theta)} \left( ia\sin\theta, 0, 1, \frac{i}{\sin\theta} \right) \,,
\end{align}
where as is typical in Kerr,
$\Delta=r^{2}-2Mr+a^{2}=(r-r_{+})(r-r_{-})$, and
$\Sigma=r^{2}+a^{2}\cos^{2}\theta$.  The roots
$r_{\pm} = M \pm \sqrt{M^{2}-a^{2}}$ are the locations of the outer
and inner horizons.  This tetrad is clearly regular at the future
horizon, where $\ell^{a}$ coincides with the horizon generator, which
in terms of the Killing vectors $\pd_{v}$ and $\pd_{\tilde{\phi}}$
and angular velocity of the horizon $\Omega_{H}$ is
\begin{align}
  \ell^{a}\Big|_{\mathcal{H}^{+}} = \frac{\pd}{\pd v} + \Omega_{H} \frac{\pd}{\pd\tilde{\phi}} \,,
  \qquad \Omega_{H} = \frac{a}{2Mr_{+}} \,.
\end{align}
The coordinate $v$ here should not be confused with the affine
parameter in Eq.~\eqref{eq:focusing}.
This tetrad is related to the very common Kinnersley
tetrad~\cite{Kinnersley:1969zza}, which is not regular at
$\mathcal{H}^{+}$, by the boost
$\ell_{\HH}=\lambda\bar{\lambda}\ell_{K}$ and
$n_{\HH}=\lambda^{-1}\bar{\lambda}^{-1}n_{K}$
where $\lambda^{-2} = 2(r^{2}+a^{2})/\Delta$.  Therefore $\ell$ and $n$
are both geodesic principal null congruences.  From the tetrad we can
assemble the inverse metric
\begin{align}
  g^{ab} = -\ell^{a}n^{b}-n^{a}\ell^{b}+m^{a}\bar{m}^{b}+\bar{m}^{a}m^{b} \,,
\end{align}
or invert for the more common form~\cite{Teukolsky:2014vca},
\begin{align}
  \begin{split}
    \rmd s^2 ={}& -\left( 1 - \frac{2Mr}{\Sigma}\right)\,
    \big(\rmd v
    - a \sin^2\theta \, \rmd \tilde\phi\big)^2 \\
    &+2 \big(\rmd v - a \sin^2\theta \, \rmd \tilde\phi\big) \,
    \big(\rmd r - a \sin^2\theta \, \rmd \tilde \phi\big)
    + \Sigma\,
    \big(\rmd\theta^2+\sin^2\theta\, \rmd\tilde\phi^2\big).
  \end{split}
\end{align}
It's interesting to inspect a few of the Newman--Penrose (NP) spin
coefficients~\cite{Newman:1961qr}
$\rho, \sigma_{\np}, \epsilon_{\np}$.  We can find $\rho$ and
$\sigma_{\np}$ from the boost transformations
$\rho_{\HH} = \lambda^{1}\bar{\lambda}^{1} \rho_{K}$ and
$\sigma_{\HH} = \lambda^{3}\bar{\lambda}^{-1} \sigma_{K}$ (where
$\{1,1\}$ and $\{3,-1\}$ are the GHP weights~\cite{Geroch:1973am} for
the spin coefficients $\rho$ and $\sigma_{\np}$),
\begin{align}
  \rho &= -m^{b} \bar{m}^{c} \cd_{c} \ell_{b} = \frac{-1}{r-ia\cos\theta} \frac{\Delta}{2(r^{2}+a^{2})} \,,\\
  \sigma_{\np} &= -m^{b} m^{c} \cd_{c}\ell_{b} = 0 \,,\\
  \epsilon_{\np} &= -\frac{1}{2} \left(
    n^{b} \ell^{c} \cd_{c}\ell_{b} -\bar{m}^{b}\ell^{c}\cd_{c}m_{b}
  \right) = \frac{M(r^{2}-a^{2})}{2(r^{2}+a^{2})^{2}} \,.
\end{align}
These are not the only non-vanishing spin coefficients, but the only
ones needed below.  The inaffinity $\kappa$ (not to be confused with
the NP coefficient $\kappa_{\np}$) and the tensor $\hat{B}_{ab}$ associated to the
congruence $\ell^{a}$ can be expressed in terms of the above spin
coefficients.  Notice that we can get the inaffinity from
\begin{align}
  \epsilon_{\np} + \bar{\epsilon}_{\np} = -n^{b}\ell^{c}\cd_{c}\ell_{b} = \kappa \,.
\end{align}
Evaluating at the horizon, we get the surface gravity of the Kerr
black hole,
\begin{align}
  \kappa_{+} = \frac{r_{+}-M}{2Mr_{+}} \,.
\end{align}
When constructing $\hat{B}_{ab}$, the projector in NP language becomes
$\gamma_{ab} = m_{a}\bar{m}_{b}+\bar{m}_{a}m_{b}$.  Assembling the
expansion, shear, and twist, we get the NP translations
\begin{align}
  \theta &= \gamma^{ab} B_{ab} = -(\rho+\bar{\rho}) \,,\\
  \omega_{ab} &= \hat{B}_{[ab]} = \frac{1}{2}(\rho-\bar{\rho})(m_{a}\bar{m}_{b}-\bar{m}_{a}m_{b}) \,,\\
  \sigma_{ab} &= \hat{B}_{\langle ab\rangle} = -m_{a}m_{b} \bar{\sigma}_{\np} - \bar{m}_{a}\bar{m}_{b} \sigma_{\np} \,.
\end{align}
Since $\sigma_{\np}=0$, indeed $\ell^{a}$ is shear-free.  Noting that
$\rho \propto \Delta$, we see that $\rho|_{\mathcal{H}^{+}}=0$, so
both the expansion and twist vanish at the future horizon, as they
must for $\ell$ to be the horizon generator.  [Contrast this with the
Kinnersley tetrad, where $\rho_{K}=-(r-ia\cos\theta)^{-1}$,
$\rho_{K}|_{\mathcal{H}^{+}}\neq 0$; and so $\ell_{K}$ has non-zero
expansion and twist at the horizon.]
Since $\ell_{\HH}$ is a null geodesic that generates
$\mathcal{H}^{+}$,
we can satisfy the conditions of the theorem with a perturbation
$h_{ab}$ which vanishes in the distant past or future, and which is
Ricci-flat to first order, $R_{ab}=\mathcal{O}(\varepsilon^{2})$.
Then the ingoing radiation gauge $\ell_{\HH}^{a}h_{ab} = 0 = h$ will
be horizon-locking.

\ack
I would like to thank
Alex Lupsasca and Eric Poisson for valuable discussions. %
This work was supported by NSF CAREER Award PHY--2047382 and
a Sloan Foundation Research Fellowship.

\section*{References}
\bibliographystyle{iopart_num}
\bibliography{notes-biblio}

\end{document}